\documentclass[11pt]{article}
\usepackage{amscd,amssymb}
\usepackage{amsmath}

\usepackage{graphicx}
\usepackage{times}

\usepackage{color}

\begin{document}


\title{Exact and simple results\\
 for the XYZ and strongly interacting fermion chains}

\bigskip\bigskip

\author{Paul Fendley and Christian Hagendorf
\medskip\\
Department of Physics, University of Virginia,
Charlottesville, VA 22904-4714 USA}

\date{August 8, 2010}

\maketitle

\begin{abstract}

We conjecture exact and simple formulas for some physical quantities in two
quantum chains. A classic result of this type is Onsager, Kaufman and
Yang's formula for the spontaneous magnetization in the Ising model,
subsequently generalized to the chiral Potts models. We conjecture
that analogous results occur in the XYZ chain when the couplings obey
$J_xJ_y + J_yJ_z + J_x J_z=0$, and in a related fermion chain with
strong interactions and supersymmetry. We find exact formulas for the
magnetization and gap in the former, and the staggered density in the latter,
by exploiting the fact that certain quantities are independent of
finite-size effects.

\end{abstract}

\bigskip

Onsager's computation of the exact partition function of the
two-dimensional Ising model \cite{Onsager1} is one of the great
triumphs of theoretical physics. This result now can be reproduced
easily, by using Kaufman's mapping of the spins to free fermions
\cite{Kaufman}. The computation of the spontaneous magnetization, by
Onsager and Kaufman \cite{Onsager} and by Yang \cite{Yang}, is a
second triumph: because the map from spins to fermions is non-local,
the computation was and remains quite intricate \cite{McCoyWu}. Their
final result is exceptionally simple.  The spontaneous magnetization
in the ordered phase $k<1$ is exactly $(1-k^2)^{1/8}$ in the
large-lattice limit; $1/k=\sinh(2J/k_BT)\sinh(2J'/k_BT)$, where $J$
and $J'$ are the usual Ising couplings for the horizontal and vertical
links of the square lattice.

It is natural to guess that the simplicity of this formula is a
consequence of the model's underlying free-fermion nature. Thus it is
remarkable that an elegant generalization of Onsager, Kaufman and
Yang's formula occurs in a series of models most decidedly not free
fermions.  The chiral Potts model is a parity-breaking $\mathbb{Z}_N$
generalization of the Ising model with some amazing properties
\cite{Howes,Gehlen}. One is that the order parameters for
spontaneously breaking the ${\mathbb Z}_N$ symmetry are given by
a formula just like the Ising model, as conjectured in ref.\
\cite{Albertini} and proved more than 15 years later in a {\it
tour de force} of Baxter's \cite{BaxterCP}. Labeling the spin at site
$j$ by a variable $\sigma_j=0\dots N-1$, the exact result as the
number of sites goes to infinity is
\begin{equation}
\langle e^{2\pi i r \sigma_j/N}\rangle=
(1-k^2)^{r(N-r)/(2N^2)}
\label{CPmag}
\end{equation}
The lattice parameter $k$ in (\ref{CPmag}) is {\em not} renormalized:
it is a coefficient of one of the terms in the corresponding
quantum Hamiltonian. Nevertheless, the expression for the order
parameters in (\ref{CPmag}) is exact for any value of $k$, ranging
from the critical point $k=1$ to the completely ordered point
$k=0$. This is unusual even for integrable models; when order
parameters can be computed they are typically given by elaborate
combinations of elliptic theta functions (see ref.\
\cite{Baxbook}).

In this paper we conjecture exact formulas analogous to (\ref{CPmag})
in two quantum chains with strong interactions: the XYZ chain along a
special line of couplings \cite{BaxXYZ,McCoy,Bazhanov,RSnew}, and
interacting fermions with supersymmetry \cite{FSd,FNS}. The
conjectures result from studying series expansions around a trivially
solvable limit, the analog of $k=0$ above. We find that for a system
with $L$ sites, the terms in these expansions up to order $L$ are {\em
independent} of $L$. We refer to such quantities as {\em scale
free}. We can thus compute them exactly by finding the ground states
explicitly for small systems. The analogs of
(\ref{CPmag}) then are obtained by summing the series. Since
this yields the correct critical exponents for the model, this
provides strong evidence that the conjecture is exact in the
$L\to\infty$ limit.

To motivate our study, we note two special properties of the chiral
Potts model.
One is that along a line in parameter space, it possesses a
useful symmetry algebra, known as the Onsager algebra
\cite{Onsager1,Gehlen}, which allows the explicit construction of an
infinite sequence of conserved quantities. A
second (under-appreciated) property is that in the
corresponding field theory in the scaling limit, the coefficient of
the Lorentz-symmetry breaking perturbation does not renormalize
\cite{CardyCP}.

Supersymmetric field theories also possess such special
properties. Because the Hamiltonian is part of the supersymmetry
algebra, supersymmetry does much more than just grouping of states
into multiplets. One can often prove the existence of zero-energy
ground states by computing the Witten index \cite{Witten}. Moreover,
in some cases there are {\em non-renormalization theorems}. For
example, in the scaling limit of the models described below, the
superpotential does not receive any corrections beyond tree level in
perturbation theory \cite{MVW}. This means that some physical
quantities (for example, the gaps of certain kink states) depend
simply on the parameters in the Hamiltonian.

This motivates us to study quantum chains whose scaling limits are
described by supersymmetric field theories. Our first example is a
special case of the well-known XYZ chain\cite{Baxbook}. The
Hilbert space $(\mathbb C^2)^{\otimes L}$ is a two-state system at
each site on the chain, and the Hamiltonian is 
\begin{equation}
H=-\sum_{j=1}^{L}\left[J_x \sigma^x_j\sigma^x_{j+1} + J_y
  \sigma^y_j\sigma^y_{j+1}  + J_z \sigma^z_j\sigma^z_{j+1} + E_0\right] 
\label{HXYZ}
\end{equation}
where the $\sigma^a$ are the Pauli matrices and $E_0$ is a
constant. For now we take periodic boundary conditions, so that
$\sigma^a_{L+1}\equiv \sigma^a_1$. When
$J_x=J_y$, the Hamiltonian preserves the numbers of up spins and down
spins individually; elsewhere these numbers are only conserved mod
2. For $L$ odd, all states including the ground state are therefore
paired by flipping all the spins. If one of the $J_a$ vanishes, the
chain can be mapped onto free fermions by the usual Jordan-Wigner
transformation; otherwise, the mapping gives interacting
fermions. Whenever $J_x=J_y$ and $|J_z|\le J_x$ (and values related by
permuting the $J_a$), the model is critical, and is called the XXZ
chain.  Along this critical line, a free-boson field
theory describes the scaling limit. Near this critical line, it can be
described by the sine-Gordon field theory. 

The field theory of the XYZ chain is supersymmetric along a particular
line in its two-parameter space (see e.g.\ ref.\ \cite{FI}). Because
the chain is integrable \cite{Baxbook}, it is easy to identify the
supersymmetric critical point in the XXZ chain: it is at
$J_z=-|J_x|/2$. The XXZ chain here has many fascinating properties
(see e.g.\ ref.\ \cite{RS}). In fact, long ago Baxter found a simple
formula for the exact ground-state energy as $L\to\infty$ along the
entire line
\begin{equation}
  J_xJ_y + J_xJ_z+J_yJ_z=0.
  \label{susyline}
\end{equation}
Namely, for $E_0=-( J_x + J_y + J_z)$, the ground-state energy
along this line goes to zero as $L\to\infty$. Moreover, it was
conjectured that the lowest eigenvalue of $H_{\rm XYZ}$ along this
line is {\em exactly} zero when $L$ is {\em odd} \cite{RS}, just as in
supersymmetric models. This was subsequently proved (for $L$ odd as
well as for $L$ even with twisted boundary conditions) at the critical
point by showing the XXZ chain has a hidden
supersymmetry relating chains with different numbers of sites
\cite{FNS,YF}. Moreover, there are a host of other fascinating and special
results along this line \cite{McCoy,Bazhanov,RSnew}, all reminiscent of the
special results occurring in fermion chains with an explicit
supersymmetry \cite{FSd,FNS,Beccaria}.
Thus the XYZ chain along the line (\ref{Jxi}) indeed should correspond
to a supersymmetric field theory in the scaling limit; for this reason
we dub this the sXYZ chain.

Our second chain is a staggered version of a fermion chain with a
built-in supersymmetry \cite{FSd,FNS}. These models are defined from
the supersymmetry operator $Q$ obeying $Q^2=0$. The
Hamiltonian $H=QQ^\dagger + Q^\dagger Q$ the commutes with
 $Q$ and $Q^\dagger$. We study the supersymmetric Hamiltonian acting
on a Hilbert space spanned by
spinless fermions, with the
additional restriction that fermions may not be on adjacent sites.
The supersymmetry operator in terms of fermion creation operators
$c_j^\dagger$ is
\begin{equation}
Q=\sum_{j}\lambda_j (1-n_{j-1})(1-n_{j+1})c_j, 
\label{Qdef}
\end{equation}
where $n_j=c^\dagger_{j}c_{j}$.
$Q$ squares to 0 for any choice of the complex
numbers $\lambda_j$, so the Hamiltonian
\begin{equation}
H_{\rm ssF}=\sum_{j=1}^{3f}  
\left[(1-n_{j-1})
(\lambda^*_j\lambda_{j+1}c^\dagger_j c_{j+1} +h.c.)(1-n_{j+2})
\ +\ |\lambda_j|^2(1-n_{j-1})(1-n_{j+1})\right]
\label{HssF}  
\end{equation}
is supersymmetric. The first term allows hopping preserving the
restriction, and the second is comprised of a chemical potential and a
next-nearest-neighbor repulsion. When the number of fermions is $f$ and the
number of sites is $3f$, for periodic boundary conditions there are
two ground states for any values of the $\lambda_j$ \cite{FSd}.  Here
we consider the staggering $\lambda_{3i}=\lambda_{3i+1}=1$ and
$\lambda_{3i+2}=z$, and so we label this model ssF (for supersymmetric
staggered Fermions). The Bethe equations for the unstaggered case
$z=1$ and for the critical sXXZ chain are the same up to boundary
conditions \cite{FNS}, and so the critical field theories must be the
same. Staggering the model perturbs it away from this critical point,
and since there is only one Lorentz-invariant supersymmetry-preserving
perturbation, its scaling limit should be the same as the
supersymmetric field theory describing sXYZ.

For the remainder of this paper we describe some of the remarkable
properties of these models. The key to much of our analysis is to
expand various quantities around a limit where the model can be solved
trivially. An amazing property of these chains is that for certain
quantities, the coefficients of the terms in this expansion are
scale free.

We parametrize the sXYZ line (\ref{susyline}) by
\begin{equation}
J_x=2s(s-3),\qquad J_y=2s(s+3),\qquad J_z=9-s^2,
\label{Jxi}
\end{equation}
so that $E_0=3(s^2+3)$. The critical points are at $s=\pm 1,\infty$,
while at the trivially solvable points $s=0,\pm 3$ only one of the
three terms in (\ref{HXYZ}) remains. At $s=0$, only $J_z\ne 0$, so the
ground state $|0\rangle$ has all spins the same.  The spontaneous
magnetization per site $M_L(s)\equiv\langle 0|\sigma^z_j |0\rangle$,
obeys $M_L(0)=1$ in the sector with an even number of down spins.  We
find, by using Maple to compute the exact ground state,
that the power-series expansions of the magnetization  for odd $L$ are
\begin{eqnarray*}
M_5&=&1-4\,{\tilde{s}}^{2}-12\,{\tilde{s}}^{4}+188\,{\tilde{s}}^{6}-844\,{\tilde{s}}^{8}+380\,{\tilde{s}}^{10}+\dots\\
M_7&=&1-4\,{\tilde{s}}^{2}-12\,{\tilde{s}}^{4}-52\,{\tilde{s}}^{6}+2516\,{\tilde{s}}^{8}-18004\,{\tilde{s}}^{10}+\dots\\
M_9&=&
1-4\,{\tilde{s}}^{2}-12\,{\tilde{s}}^{4}-52\,{\tilde{s}}^{6}-284\,{\tilde{s}}^{8}+33516\,{\tilde{s}}^{10}+\dots\\
M_{11}&=&
1-4\,{\tilde{s}}^{2}-12\,{\tilde{s}}^{4}-52\,{\tilde{s}}^{6}-284\,{\tilde{s}}^{8}-1764\,{\tilde{s}}^{10}+\dots
\end{eqnarray*}
where $\tilde{s}=s/3$.
The trend is obvious: the order $s^{n}$ terms in the
expansion are independent of $L$ when $n<L$. The magnetization appears
to be scale free near $s=0$. Doing this to $L=17$ yields what is
presumably the exact expansion as $L\to\infty$:
$$M_{L}(s)=
1-4\,{\tilde{s}}^{2}-12\,{\tilde{s}}^{4}-52\,{\tilde{s}}^{6}-284\,
{\tilde{s}}^{8}-1764\,{\tilde{s}}^{10}
-11820 {\tilde{s}}^{12} - 83220 {\tilde{s}}^{14}
- 606780 {\tilde{s}}^{16} +\dots
$$

To understand how to sum the series and find a simple formula
$M_\infty(s)$, we examine the expected behavior at the critical point
$s{=}1$. The dimension of the ``thermal'' operator that perturbs away
from $s{=}1$ onto the sXYZ line is $4/3$, while the dimension of the
magnetization operator is expected to be $1/3$ \cite{Baxbook}.
Indeed, the finite-size values at criticality fit nicely to
$M_L(1)\approx .95527\, L^{-1/3}(1 +O(L^{-2}))$. Thus as $s\to 1^-$,
$M_\infty(s)$ should vanish as $(1-s)^\beta$ with
$\beta=(1/3)/(2-4/3)=1/2$.
This square-root singularity suggests looking at
the series expansion of $(M_L(s))^2$:
$$(M_L(s))^2=1-8 \tilde{s}^2-8 \tilde{s}^4-8 \tilde{s}^6-8 \tilde{s}^8-
8 \tilde{s}^{10} 
- 8 {\tilde{s}}^{12} - 8 {\tilde{s}}^{14}
- 8 {\tilde{s}}^{16} 
+O(s^{L+1})\ .$$
Summing this series yields a conjecture for the exact
magnetization in the ordered phase $s<1$:
\begin{equation}
M_\infty(s)= 3\left(\frac{1-s^2}{9-s^2}\right)^{1/2}\ .
\label{Mconj}
\end{equation}
We emphasize that we do not assume anything about the behavior
at the critical point $s=1$; the only role of the
scaling argument is to suggest that we square $M$. The fact that the
expected critical behavior for the magnetization emerges from the
expansion around $s=0$ is to us a compelling argument that the
formula (\ref{Mconj}) is exact.

\begin{figure}[ht] 
\begin{center} 
\includegraphics[width= .6\textwidth]{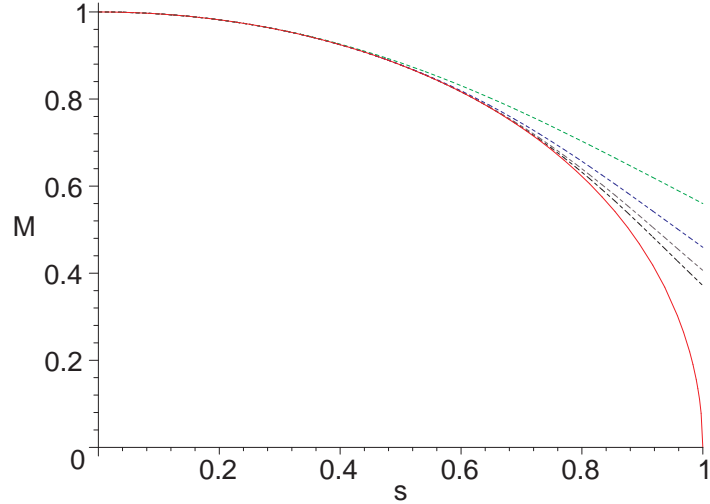} 
\caption{$M_L(s)$ for $s\le 1$; the solid red curve is the conjecture
  for $M_\infty(s)$, while the dashed curves
  are for $L=5,9,13,17$.}
\label{fig:M} 
\end{center} 
\end{figure}

We plot this function and the finite-size curves in figure
\ref{fig:M}. Even with the finite-size effects near $s=1$,
it is clear that the finite-$L$ curves are approaching the conjectured
curve. A numerical calculation using the iTEBD method gives the same
curve to high accuracy \cite{Pollmann}. Moreover, with a change of
variables, \eqref{Mconj} gives the same ``$q$-series'' obtained by
exploiting the integrability of the chain. The corresponding quantity
in the eight-vertex model, the spontaneous polarization, is
\cite{baxterkelland,jimbo}
\begin{equation}
 P_0=\frac{4\pi}{2\pi-\eta}
\left(\frac{\vartheta _2\left(0,q^{1/(2-4\eta/\pi)}\right)}
{\vartheta _2\left(0,q^{1/2}\right)}\right)^2
 \label{bkformula}
\end{equation}
when written in terms of Jacobi theta functions \cite{whittaker}. The
sXYZ line \eqref{susyline} corresponds to setting the crossing
parameter $\eta=\pi/3$. Using the Boltzmann weights of the
eight-vertex model at this point, we find
\begin{equation}
  s =
  3\left(\frac{\vartheta_2(\pi/3,q^{1/2})}{\vartheta_1(\pi/3,q^{1/2})}
\right)^2 
\label{sq}
\end{equation}
for $0<s<1$. Inserting this into \eqref{Mconj} yields
the same expansion in $q$ as that of \eqref{bkformula} to $\sim 500$ terms.

We have found other exact formulas using these methods. Letting $H^a_j
= \sigma^a_j\sigma^a_{j+1}$, for $s<1$
\begin{eqnarray}
\langle 0|H^z_j|0\rangle
&=&1+4\tilde{s}^2(-1+\tilde{s}^2+3\tilde{s}^4+5\tilde{s}^6+7\tilde{s}^8+\dots)\nonumber\\
&=&1+12\frac{s^2(s^2-3)}{(s^2-9)^2} + O(s^{L+1})\label{hzdisordered}
\end{eqnarray}
We can find this expectation value for $s>1$ by expanding around
$s=3$, where $J_x=J_z=0$. 
Letting $t=(3-s)/6$,
\begin{eqnarray}
\langle 0|H^z_j|0\rangle
&=&\frac{1}{2}\left(2t+3t^2+4t^3+5t^4+\dots\right)\nonumber\\
&=&\frac{(s+9)(3-s)}{2(3+s)^2}\ +\  O(t^{L+1})\ .
\label{hzferro}
\end{eqnarray}
Using the $q$-series representation \eqref{sq} we showed analytically
that \eqref{hzdisordered} matches the exact results \cite{boos}.  The
expectation values of $H^x_j$ and $H^y_j$ can be found from these by
using the duality symmetries $s\to (3-s)/(s+1)$ and $s\to -s$, or by
using Hellmann-Feynman theorem $\langle 0|\rm{d}H_{\rm
sXYZ}/\rm{d}s|0\rangle =0$.

%

Not only ground-state properties are scale free: the gap is as well,
and obeys an elementary formula. To define the gap, we exploit the
fact that there is a spontaneously broken ${\mathbb Z}_2$ symmetry
away from the critical points. It is thus natural to think of the
gapped excited states as kinks separating regions of the two ground
states. This picture is supported by the computation of the exact
scattering matrix for these kinks in the supersymmetric field theory
\cite{FI}. For an odd number of sites and periodic boundary
conditions, we expect the lowest-energy excited states to be two-kink
states. Since the kinks interact, the energy is less than twice the
kink gap. Thus to define the gap to the one-kink state, we consider an
{\em even} number of sites with twisted boundary conditions (a
spin-flip defect): $\sigma^z_{n+1}=-\sigma^z_1$,
$\sigma^y_{n+1}=-\sigma^y_1$, and $\sigma^x_{n+1}=+\sigma^x_1$.  Near
$s=0$, the interactions away from the boundary favor aligning the
spins, but the twist forces the energy to be order $J_z$.

The result for the gap is more transparent when we rescale the
Hamiltonian $H\to H/s^2$, and consider the region between the
trivially solvable point $s=3$ and the critical point at $s\to\infty$;
the gap in other regions is obtained by exploiting duality.
We found the exact one-kink energy $\Delta$ for
sizes up to $L=10$. Expanding this in a power series around $s=3$ in
terms of $v=1-3/s$, we find
\begin{eqnarray}
\nonumber
\Delta_L&=& 4-6\,v+3\,v^2/2+v^3/4+3\,v^4/32+\dots\\
&=& 4\left(\frac{3}{s}\right)^{3/2} + O(v^{L/2})
\end{eqnarray}
Thus at the critical point $s\to\infty$, the gap vanishes with
exponent $\nu=3/2$. This is exactly what one expects with
dimension-4/3 thermal operator: $\nu=1/(2-4/3)=3/2$. 

We now turn to the supersymmetric staggered fermion model with
Hamiltonian (\ref{HssF}), and show that not only does it possess
scale-free quantities similar to those of the sXYZ chain, but that the
models are deeply related on the lattice, not just in their scaling
limits.  For $f$ fermions on $3f$ sites, $H_{\rm ssF}$ has two
zero-energy ground states like $H_{\rm sXYZ}$. Here, however, the two
ground states are not related by symmetry: because the fermions cannot
occupy adjacent sites, there is no analog of spin-flip symmetry. To
define basis vectors for the two-dimensional space of zero-energy
states unambiguously, we exploit the parity symmetry $j\to
3f+1-j$. One ground state, denoted $|+\rangle_z$, is even under
parity, while the other ground state $|-\rangle_z$ is odd.  These two
ground states are quite different from each other, as is easy to see
by studying them in the solvable limits $z\to 0,\infty$. Since
$H=QQ^\dagger +Q^\dagger Q$, any zero-energy ground state must be
annihilated by both $Q$ and $Q^\dagger$.  Letting $|j\rangle$ label
the three states with a fermion on every third site $3i+j$, we have at
$z\to\infty$, $|\pm \rangle_\infty=(|1\rangle \pm
|3\rangle)/\sqrt{2}$. In the other limit, $|+\rangle_0=|2\rangle$, but
the odd-parity ground state is a sum over all configurations without a
fermion on the sites $3i+2$: $|-\rangle_0 = \prod_{i=1}^{f}
(c^\dagger_{3i+1}-c^\dagger_{3i}) |{\rm empty}\rangle/2^{f/2}\ .  $

\begin{figure}[ht] 
\begin{center} 
\includegraphics[width= .6\textwidth]{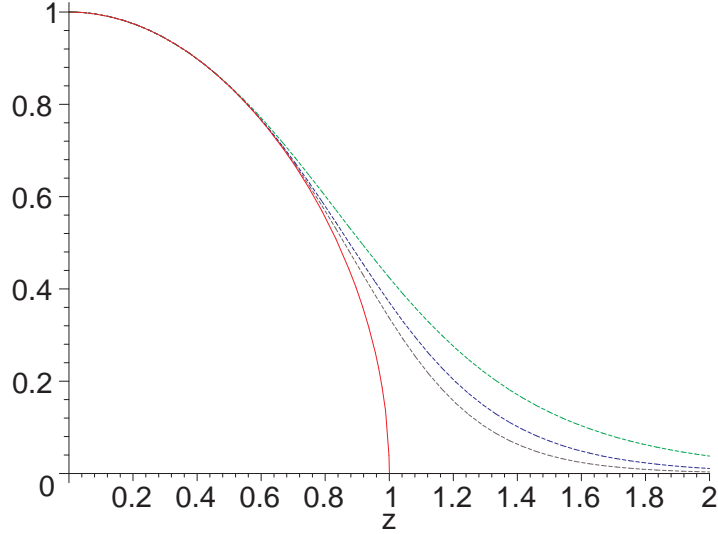} 
\caption{$D^+(z) - D^-(z)$; the solid red curve is conjecture
  (\ref{Dm}), and the dashed curves are for sizes $3f=12,18,24$.}
\label{fig:Nlow} 
\end{center} 
\end{figure}

We find exact formulas for the staggered fermion densities $D^\pm (z)=
{}_z\langle \pm | c_{3i-1}^\dagger c_{3i-1} |\pm \rangle_z$. These
have been studied at the critical point $z=1$, using numerics
\cite{Beccaria} and using conformal field theory \cite{Liza}, and we
extend these results to all $z$. In the solvable limits, we have
$D^+(\infty)=D^-(\infty)=D^-(0)=0$ and $D^+(0)=1$. Moreover, at the
critical point $z=1$, the full translation symmetry of the model is
restored. This requires that $D^+(1)+D^-(1)=2/3$. We thus expect that
$D^+ - D^-$ behaves like the magnetization in the sXYZ chain,
vanishing as $f\to\infty$ when $z\ge 1$, but non-zero for $z<1$. By
finding the exact ground state in sizes up to $f=8$ (24 sites), we
obtain for $z$ small
\begin{eqnarray}
\nonumber
D^+ + D^- &=&
1-3\tilde{z}^{2}+{3}\tilde{z}^{4}-
{3}\tilde{z}^{6}+ {3}\tilde{z}^{8}-\dots\\
&=& \frac{8-2z^2}{8+z^2} + O(z^{2f}),\\ 
\nonumber
D^+ - D^- &=&
1-{5}\tilde{z}^{2}-{3}\tilde{z}^{4}-{29}\tilde{z
}^{6}-{131}\tilde{z}^{8}-\dots\\
&=&\frac{8\sqrt{1-z^2}}{8+z^2}+ O(z^{2f})
\label{Dm}
\end{eqnarray}
where $\tilde{z}=z/\sqrt{8}$. 
We see the same square-root singularity in $D^+ -
D^-$ that we did for the magnetization in sXYZ.
The series
expansion around $z=\infty$ gives
\begin{eqnarray}
\nonumber
D^+ + D^- &=&
\frac{2}{z^2}-
\frac{6}{z^4}+\frac{26}{z^6}-
\frac{134}{z^8}+\frac{762}{z^{10}}-\frac{4614}{z^{12}}+\dots\\
&=& 
\frac{4}{z^2+z\sqrt{8+z^2}+2} + O(z^{-4f})
\label{Dp}
\end{eqnarray}
The finite-size curves for $D^+ + D^-$ are almost
indistinguishable from the asymptotic form, because of the exact result
at $z=1$ and the scale-free behavior in the small- and large-$z$
limits.  The curves for $D^+ - D^-$ are plotted in figure
\ref{fig:Nlow}.

Field-theory dualities need not be exact in the corresponding
lattice models, or can be very subtle (e.g.\ the Kramers-Wannier
duality of the Ising model). Thus even though the sXYZ chain has
the duality symmetry $s\to (3-s)/(s+1)$, the corresponding duality is
not obvious in the ssF chain. Nevertheless, we have
non-trivial evidence that there is such a duality exchanging
the $|z|>1$ and $|z|<1$ phases. This becomes apparent when
we simplify (\ref{Dp}) by defining the new coupling
$S=3z/\sqrt{z^2+8}$, so that $D^+ + D^-= 2(3-S)/(3S+3)$ asymptotically
for $z>1$. The $z=1$ critical point is at $S=1$, while the solvable
points $z=0$ and $z=\infty$ correspond to $S=0$ and $S=3$
respectively, the same as the value of $s$ in the sXYZ chain.  We find
that
$$D^+(S)\,D^-(S) = D^+(\widehat{S})\, D^-(\widehat{S}) $$ for
$\widehat{S}=(3-S)/(S+1)$. This relation holds for all finite sizes
up to $3f=24$ sites, and of course for the
asymptotic formulas as well. It thus seems very likely that this is a
general symmetry of the ssF chain, but we have not yet found the
corresponding symmetry of the Hamiltonian.

The relation between the sXYZ and ssF chains goes even deeper. In a
remarkable series of papers \cite{Bazhanov}, Bazhanov and Mangazeev
showed that (at least for small systems)
the ground states themselves are related to the tau functions of the
Hamiltonian hierarchy of the Painlev\'e VI non-linear differential
equation. They find a recursion relation for the coefficient of the
state with all spins down in the wavefunction, normalized so that it
is a polynomial in $s$. This same polynomial appears in the ground
state in our ssF chain! It appears (up to a convention-dependent
overall power of $S$) for example as the coefficient in $|+\rangle_z$
of the state $|2\rangle$ defined above, when $z$ is rewritten in terms
of $S$. Moreover, the normalizations of the ground states are related
to the same polynomials, just as in ref. \cite{Bazhanov}. Thus
the ssF ground states can be related in the same fashion to Painlev\'e
VI.

We have presented conjectures for exact results in two interacting
chains. These include simple formulas for the spontaneous
magnetization and the gap in the XYZ chain when the scaling limit is a
supersymmetric field theory. We believe that the evidence for these
conjectures is convincing. Moreover, these chains quite obviously have
a great deal of symmetry structure left to be uncovered. In
particular, all the evidence -- the scale free property, the important
role of supersymmetry, and the precise relations between the ground
states of the two chains -- makes it seem likely to us that there is
an infinite-dimensional symmetry in both models similar to the Onsager
algebra \cite{Onsager1}. Each model will then correspond to a
different presentation of this algebra.

\medskip

This work was supported by the NSF grant DMR/MSPA-0704666.


\end{document}